\documentclass[apj]{emulateapj}
\usepackage{apjfonts,mathptmx}

\begin{document}

\def\agile {\emph{AGILE}}
\def\xmm {\emph{XMM-Newton}}
\def\cha {\emph{Chandra}}
\def\flux {\mbox{erg cm$^{-2}$ s$^{-1}$}}
\def\lum {\mbox{erg s$^{-1}$}}
\def\nh {$N_{\rm H}$}

\title{DISCOVERY OF NEW GAMMA-RAY PULSARS WITH \emph{AGILE}}

\shorttitle{New Gamma-ray Pulsars}
\shortauthors{A.~Pellizzoni et al.}
\journalinfo{This is an author-created, un-copyedited version of an article accepted for publication in The Astrophysical Journal Letters. IOP Publishing Ltd is not responsible for any errors or omissions in this version of the manuscript or any version derived from it. The definitive publisher authenticated version is available online at 10.1088/0004-637X/695/1/L115.}
 \submitted{Received 2009 January 08; Accepted 2009 February 27}

\author{A.~Pellizzoni\altaffilmark{1}, M.~Pilia\altaffilmark{1,2,3}, 
A.~Possenti\altaffilmark{1}, A.~Chen\altaffilmark{2,4}, 
A.~Giuliani\altaffilmark{2}, A.~Trois\altaffilmark{5}, 
P.~Caraveo\altaffilmark{2}, E.~Del Monte\altaffilmark{5}, 
F.~Fornari\altaffilmark{2}, F.~Fuschino\altaffilmark{6}, 
S.~Mereghetti\altaffilmark{2}, M.~Tavani\altaffilmark{5,7}, 
A.~Argan\altaffilmark{5}, M.~Burgay\altaffilmark{1}, 
 I.~Cognard\altaffilmark{8}, 
A.~Corongiu\altaffilmark{8}, E.~Costa\altaffilmark{5},
N.~D'Amico\altaffilmark{1},
A.~De Luca\altaffilmark{2,9,10}, P.~Esposito\altaffilmark{2,9}, 
Y.~Evangelista\altaffilmark{5}, M.~Feroci\altaffilmark{5}, 
S.~Johnston\altaffilmark{11}, M.~Kramer\altaffilmark{12}, 
F.~Longo\altaffilmark{13}, M.~Marisaldi\altaffilmark{6}, 
G.~Theureau\altaffilmark{8,14},
P.~Weltevrede\altaffilmark{11}, G.~Barbiellini\altaffilmark{13}, 
F.~Boffelli\altaffilmark{9,15}, A.~Bulgarelli\altaffilmark{6}, 
P. W. Cattaneo\altaffilmark{9}, V.~Cocco\altaffilmark{5}, 
F.~D'Ammando\altaffilmark{5,7}, G.~De Paris\altaffilmark{5}, 
G.~Di Cocco\altaffilmark{6}, I.~Donnarumma\altaffilmark{5}, 
M.~Fiorini\altaffilmark{2}, T.~Froysland\altaffilmark{4,7}, 
M.~Galli\altaffilmark{16}, F.~Gianotti\altaffilmark{6}, 
C.~Labanti\altaffilmark{6}, I.~Lapshov\altaffilmark{5}, 
F.~Lazzarotto\altaffilmark{5}, P.~Lipari\altaffilmark{17}, 
T.~Mineo\altaffilmark{18}, A.~Morselli\altaffilmark{19}, 
L.~Pacciani\altaffilmark{5}, F.~Perotti\altaffilmark{2}, 
G.~Piano\altaffilmark{5,7},
P.~Picozza\altaffilmark{19}, M.~Prest\altaffilmark{3}, 
G.~Pucella\altaffilmark{5}, M.~Rapisarda\altaffilmark{20}, 
A. Rappoldi\altaffilmark{9},
S.~Sabatini\altaffilmark{5,17},
 P.~Soffitta\altaffilmark{5}, 
M.~Trifoglio\altaffilmark{6}, E.~Vallazza\altaffilmark{13}, 
S.~Vercellone\altaffilmark{2}, V.~Vittorini\altaffilmark{7}, 
A.~Zambra\altaffilmark{2}, D.~Zanello\altaffilmark{17}, 
C.~Pittori\altaffilmark{21}, F.~Verrecchia\altaffilmark{21}, 
B.~Preger\altaffilmark{21}, P.~Santolamazza\altaffilmark{21}, 
P.~Giommi\altaffilmark{21}, L.~Salotti\altaffilmark{22}, and
G.~F.~Bignami\altaffilmark{10}}

\affil{$^{1}$ INAF--Osservatorio Astronomico di Cagliari, localit\`a Poggio dei Pini, strada 54, I-09012 Capoterra, Italy; \url{apellizz@ca.astro.it}}
\affil{$^{2}$ INAF/IASF--Milano, Via E.~Bassini 15, I-20133 Milano, Italy}
\affil{$^{3}$ Dipartimento di Fisica, Universit\`a dell'Insubria, Via Valleggio 11, I-22100 Como, Italy}
\affil{$^{4}$ CIFS--Torino, Viale Settimio Severo 3, I-10133, Torino, Italy}
\affil{$^{5}$ INAF/IASF--Roma, Via del Fosso del Cavaliere 100, I-00133 Roma, Italy}
\affil{$^{6}$ INAF/IASF--Bologna, Via Gobetti 101, I-40129 Bologna, Italy}
\affil{$^{7}$ Dipartimento di Fisica, Universit\`a ``Tor Vergata'', Via della Ricerca Scientifica 1, I-00133 Roma, Italy}
\affil{$^{8}$ Laboratoire de Physique et Chimie de l'Environnement, CNRS, F-45071 Orleans, France}
\affil{$^{9}$ INFN--Pavia, Via Bassi 6, I-27100 Pavia, Italy}
\affil{$^{10}$ Istituto Universitario di Studi Superiori, V.le Lungo Ticino Sforza 56, 27100 Pavia, Italy}
\affil{$^{11}$ Australia Telescope National Facility, CSIRO, P.O.~Box~76, Epping NSW~1710, Australia}
\affil{$^{12}$ University of Manchester, Jodrell Bank Observatory, Macclesfield, Cheshire SK11 9DL, UK}
\affil{$^{13}$ Dipartimento di Fisica, Universit\`a di Trieste and INFN--Trieste, Via Valerio 2, I-34127 Trieste, Italy}
\affil{$^{14}$ GEPI, Observatoire de Paris, CNRS, Universit\'e Paris Diderot; Place Jules
Janssen F-92190 Meudon, France}
\affil{$^{15}$ Dipartimento di Fisica Nucleare e Teorica, Universit\`a di Pavia, Via Bassi 6, Pavia, I-27100, Italy}
\affil{$^{16}$ ENEA--Bologna, Via Biancafarina 2521, I-40059 Medicina (BO), Italy}
\affil{$^{17}$ INFN--Roma ``La Sapienza'', Piazzale A. Moro 2, I-00185 Roma, Italy}
\affil{$^{18}$ INAF/IASF--Palermo via U. La Malfa 153, I-90146, Palermo, Italy}
\affil{$^{19}$ INFN--Roma ``Tor Vergata'', Via della Ricerca Scientifica 1, I-00133 Roma, Italy}
\affil{$^{20}$ ENEA--Roma, Via E. Fermi 45, I-00044 Frascati (Roma), Italy}
\affil{$^{21}$ ASI--ASDC, Via G. Galilei, I-00044 Frascati (Roma), Italy}
\affil{$^{22}$ ASI, Viale Liegi 26 , I-00198 Roma, Italy}

\begin{abstract}


Using gamma-ray data collected by the \emph{Astro-rivelatore Gamma ad Immagini LEggero} (\emph{AGILE}) satellite over
a period of almost one year (from 2007 July to 2008 June), we
searched for pulsed signals from 35 potentially interesting radio
pulsars, ordered according to $F_{\gamma}\propto \sqrt{\dot{E}}
d^{-2}$ and for which contemporary or recent radio data were
available. \emph{AGILE} detected three new top-ranking nearby and
Vela-like pulsars with good confidence both through timing and spatial
analysis.  Among the newcomers we find  pulsars 
with very high rotational energy losses, such as 
the remarkable PSR\,B1509--58
with a magnetic field in excess of 10$^{13}$ Gauss, and
PSR\,J2229+6114 providing a reliable identification for the previously 
unidentified EGRET source 3EG\,2227+6122.
Moreover, the powerful millisecond pulsar B1821--24, in the globular
cluster M28, is detected during a fraction of the observations.
Four other promising gamma-ray pulsar candidates, among which is the notable J2043+2740 with an age 
in excess of 1 million years, show a possible detection in the
timing analysis only and deserve confirmation.

\end{abstract}

\section{Introduction}

Although the bulk of the electromagnetic energy output of spin-powered pulsars is typically 
expected above 10 MeV, only $\sim$ 0.5\% of the radio pulsar population has been clearly 
identified in the gamma-ray domain \citep{thompson04,ppp08,halpern08}. Such meager
harvest is to be ascribed, most probably, to the relatively low sensitivity of gamma-ray instruments with 
respect to radio and X-ray telescopes. 
Poor gamma-ray pulsar statistics has been a major difficulty in
assessing the dominant mechanism which channels pulsar rotational
energy into high-energy emission as well as understanding the
sites where charged particles are accelerated.

The dominant mechanisms and sites of the emission most probably
depend on the rotational period $P$ and magnetic field $B$ of the
neutron stars, with the millisecond pulsars (e.g.
\citealt{harding05}) behaving differently from the ``classical'',
higher magnetic field ones. Population synthesis simulations,
featuring comprehensive statistical analysis of diverse models of
emission geometry and beaming, yielded very different numbers of
radio-loud and radio-quiet pulsars potentially detectable as
gamma-ray emitters (see e.g., \citealt{gonthier07}). However, such
simulations are poorly constrained by the source sample currently
available and only a much larger sample of gamma-ray pulsars
could help discriminating different emission models.

\emph{Astro-rivelatore Gamma ad Immagini LEggero} (\agile) is a
scientific mission of the Italian Space Agency, dedicated to
high-energy astrophysics \citep{tavani08} launched on 2007 April
23. The sensitivity to photons with energy in the range 30
MeV to 30 GeV of the \agile\ Gamma-Ray Imaging Detector (GRID;
\citealt{prest03,barbiellini01}) together with a time tagging
accuracy of a few $\mu$s and a very large field of view
($\gtrsim$$60\degr$ radius) all make \agile\ perfectly suited for
the detection of the time signature of new gamma-ray pulsars.

The large field of view of \agile\ allows long uninterrupted observations and 
simultaneous monitoring of tens of nearby radio pulsars belonging to the ``gamma-ray 
pulsar region'' of the $P$--$\dot{P}$ diagram characterized by $B>2\times10^{11}$ G and 
spin-down energy $\dot{E}_{\rm{rot}}>1.3\times10^{33}$ erg s$^{-1}$ 
\citep{pellizzoni04agile}.


At variance with the behavior at soft X-ray energies, where the
emission is proportional to the rotational energy loss over the
squared distance factor ($F_{\rm{X}}\propto \dot{E} d^{-2}$,
where $d$ is pulsar distance; e.g. \citealt{possenti02}), the
expected gamma-ray flux of radio pulsars is directly correlated
to the Goldreich--Julian current/open field line voltage
\citep{cheng80,harding81}. It can be
estimated according to the law $F_{\gamma}\propto \sqrt{\dot{E}}
d^{-2}$ \citep{kanbach02,arons96}, which is known
to reasonably fit EGRET pulsars.
The large dispersion  of such fit
(probably due to different beaming fractions)
provides the minimum/maximum normalization values,
allowing a worst/best case approach for the gamma-ray flux estimates.
Following such an approach, we have built a sample of $\sim$ 100
radio-loud pulsars which are likely to be above the \agile\
sensitivity threshold ($F_{\rm{min}}>2\times10^{-8}$ ph~cm$^{-2}$
s$^{-1}$ at $E>100$ MeV). Of course, we are fully aware that the
actual number of detections will depend upon the emission model
geometry and efficiency.

Top-ranking targets with poor \agile\ exposure, well-known pulsars
already deeply investigated at the epoch of EGRET observations,
and sources reserved to the \agile\ Guest Observer Program were
excluded from our list which encompasses 35 gamma-ray pulsar candidates.\footnote{In ranking order:
J0737--3039*, J1833--1034, J1744--1134, J1740+1000, J1747--2958,
J2043+2740, J1730--2304, J1513--5908, J1524--5625, J1909--3744*,
J1357--6429, J1531--5610, J1809--1917, J1617--5055, J1803--2137,
J1801--2451, J0940--5428, J1549--4848, J1718--3825, J1824--2452,
J1730--3350, J0900--3144*, J1420--6048, J1739--3023, J1751--2857*,
J1804--2717*, J2229+6114, J1105--6107, J1721--2457, J1124--5916,
J1722--3712, J1740--3015, J1823--3106, J1745--3040, J1016--5857 (the asterisk indicates binary systems).
See \url{http://agile.asdc.asi.it} for details about \agile\ Data 
Policy and \agile\ Team Targets List.}
In this Letter we present results about isolated pulsars only.
The five pulsars in binary systems of our sample will be discussed in subsequent works.

Since gamma-ray pulsar detection must start from an at least
approximate knowledge of recent pulsar ephemeris, a dedicated
pulsar radio monitoring campaign has been undertaken. 
Conservatively, we present here timing analysis results and detection claims
only for pulsars having simultaneous radio ephemeris for the whole relevant
\agile\ observing epochs.
Our campaign will continue throughout the  \agile\ mission for most
of the targets (accordingly to their visibility), using two
telescopes (Jodrell Bank and Nan\c cay) of the European Pulsar
Timing Array (EPTA), as well as the Parkes radio telescope of The
Australia Telescope National Facility (ATNF).
Details about radio data analysis procedure in use for \agile\ are described in \citet{ppp08}.
Timing solutions for the detected pulsars can be requested by contacting the corresponding author.

\section{\emph{AGILE} Observations and Data analysis}

\begin{deluxetable*}{lrrrrrrrrrrrrr}
\tablewidth{43pc}
 \tablecaption{\label{newpulsars} 
Emission parameters of the seven pulsars discussed in text (Firmly detected
gamma-ray pulsars are reported in the upper part of the table)}
\tablehead{
PSR Name & G.Lon. & G.Lat. & $P$ & $\tau$\tablenotemark{a} & $D$\tablenotemark{a} & log $\dot{E}$  & $\chi^2_{red}$$(N_{st})$\tablenotemark{b} & $\sigma_{time}$\tablenotemark{c} & $\sigma_{space}$\tablenotemark{d}  & $F_\gamma$\tablenotemark{d} & log $L_\gamma$\tablenotemark{e} & $L_\gamma/\dot{E}$ \\
&           {\footnotesize deg}    &  {\footnotesize deg}   &  {\footnotesize ms}   & {\footnotesize yr} & {\footnotesize kpc} & {\footnotesize erg s$^{-1}$} & &  & & & {\footnotesize erg s$^{-1}$}
}
\startdata
J2229+6114 & 106.65 &  2.95 & 51.6 & $1.0\times10^{4}$ & 12.0  & 37.35 & 6.0(36)  & 5.0  & 7.5 & $26\pm4$ & 35.36 & 0.01 \\
J1513--5908 & 320.32 & -1.16 & 150.7 & $1.6\times10^{3}$ & 5.8  & 37.25 & 4.2(3)  & 4.0  & 6.4 & $34\pm6$\tablenotemark{f} & 35.04 & 0.006 \\ 
J1824--2452 & 7.80 & -5.58 & 3.0 & $3.0\times10^{7}$  & 4.9 & 36.35 & 4.2(1)   & 4.2   & 3.6 & $18\pm5$  & 34.62 & 0.02 \\
\hline
J1016--5857 & 284.08 & -1.88  & 107.4 &$2.1\times10^{4}$ & 9.3 & 36.41 & 6.0(69)   & 4.8   & 12.3 & $62\pm6$\tablenotemark{f} & 35.71 & 0.2 \\
J1357--6429 & 309.92 & -2.51  & 166.1 &  $7.3\times10^{3}$  & 4.1    & 36.49 &  5.2(7)   & 4.7  & 1.8 & $<$14 & $<$34.35 & $<$0.007 \\
J2043+2740 & 70.61  & -9.15 & 96.1  & $1.2\times10^{6}$   & 1.1 & 34.75  & 4.1(1)   & 4.2   & 0.6 & $<$6 & $<$32.84 & $<$0.01  \\
J1524--5625 & 323.00 &  0.35 & 78.2  &  $3.2\times10^{4}$ & 3.8 & 36.51 & 4.6(4)   & 4.3   & 1.0 & $<$16 & $<$34.34 & $<$0.007\\
\enddata
\tablenotetext{a}{Dispersion measure distance and characteristic age obtained from ATNF Pulsar Catalog (http://www.atnf.csiro.au/research/pulsar/psrcat/; \citealt{manchester05}). For J2229+6114, \citet{halpern01} derived a much shorter alternative distance of 3 kpc from X-ray absorption.}
\tablenotetext{b}{Pearson's $\chi^2$ statistics applied to the 10-bin, $E>50$ MeV folded pulse profiles
($N_{st}$ is the number of steps over period search grid, see text).} 
\tablenotetext{c}{Pulsed emission detection significance (weighted accounting for $N_{st}$).}
\tablenotetext{d}{Likelihood analysis detection significance and flux (or 2$\sigma$ upper limit)
in units of 10$^{-8}$ ph cm$^{-2}$ s$^{-1}$ ($E>100$ MeV), obtainined by search for a source near the pulsar direction.}
\tablenotetext{e}{
$E>100$ MeV luminosity assuming traditional 1 sr beam emission and photon index 2.0 (for J2229+6114 the measured photon index of 2.2 has been used).
}
\tablenotetext{f}{Possibly multiple sources. Source confusion may
affect flux and significance.}
\end{deluxetable*}
Pulsar data were collected since the early phases of the mission.
Timing observations suitable for pulsed signal analysis started
in 2007 July (at orbit 1146) after engineering tests on the
payload. In this Letter, we analyze data collected up to 2008
June 30.

\agile\ pointings consist of long exposures, typically lasting
10--30 days, slightly drifting ($\lesssim$1\degr\ day$^{-1}$) from
the starting  direction in order to match solar panel
illumination constraints. The \agile\ total exposure (computed
with the GRID scientific analysis task AG\_ExpmapGen) for the 11
month data span considered is typically $\lesssim$$10^{9}$
cm$^2$ s for each target.

Data screening, particle background filtering and event direction
and energy reconstruction were performed by the \agile\ Standard
Analysis Pipeline (BUILD-15).  We adopted the \agile\ event
extraction criteria and timing procedures calibrated and
optimized with the observations of known gamma-ray pulsars as
described in \citet{ppp08}. In particular, we performed our
timing analysis looking for pulsed signals using both $E>50$ MeV G
class events (i.e. events identified  with good confidence as
photons) and the combination of G+L events (L events are
significantly contaminated by particle background) collected
within 60\degr\ from the center of GRID field of view and with an
extraction radius around pulsar positions optimized as a function
of photon energy according to the point-spread function (PSF) and
background level of \agile. For noisy observations in confused
regions (e.g. for J2229+6114), an extraction radius of
$\sim$2\degr\ (well below $E>50$ MeV PSF\footnote{The on-axis \agile\ PSF 67\% containment radius
is $\sim$ 10\degr\ for $E=50$ MeV, $\sim$ 5\degr\ for $E=100$ MeV, and $\sim$ $0\fdg5$ for $E=1$ GeV.})
 was used to optimize the signal-to-noise ratio.
Source position accuracy is $<$$0\fdg5$ at present level of targets exposure.

Given the measured \emph{AGILE}'s time tagging accuracy of $<$200 $\mu$s and the
good radio monitoring (i.e., valid epoch range, adequate number of time
of arrivals (ToAs)) available for the majority of our targets,
the most significant pulsed 
signal detection is typically expected within the errors of the radio ephemeris 
frequency values. In particular, we performed standard epoch folding\footnote{When 
adequate radio observations covering the time span of the gamma-ray observations 
are available (i.e. where the $\it{WAVE}$ terms are included in TEMPO2 ephemeris 
files; \citep{hobbs04,hobbs06,edwards06}), we account also for the pulsar timing 
noise in the folding procedure as reported in \citet{ppp08}.}  over a frequency 
 range defined according to 3$\sigma$ errors of radio ephemeris.

The frequency resolution is 0.5/$T_{ds}$ (where $T_{ds}$ is the time span
relative to the coverage of each target) and we oversampled it by a factor of 10 
(0.05/$T_{ds}$ is the frequency step of our grid). Since the \agile\ useful data span 
for pulsar detection is typically of the order of few months, implying a period
search resolution of $<$10$^{-10}$ s, the resulting number of independent 
period search trials turns out to be few tens while no independent trials on period derivatives are needed.

Pearson's $\chi^2$ statistics is applied to the 10-bin folded pulse profiles 
resulting from each set of spin parameters, yielding the 
probabilities (weighted for the number of trials performed on the data set) 
of sampling a uniform distribution, assessing the significance of the pulsed signal
(sinusoidal or not). 
In fact, applying this statistics to the well-known EGRET pulsars \citep{ppp08},
we obtained a perfect match between the best period resulting from \agile\ gamma-ray data
and the period predicted by the radio ephemeris with discrepancies comparable to
the period search resolution.
Furthermore, we verified our timing results also applying bin-independent
parameter-free statistics as the $Z^2_n$ test \citep{buccheri83}
and the $H$-test \citep{dejager89} that are typically more sensitive than
$\chi^2$ tests for the search of sinusoidal pulses.
In particular, the use of the $H$-test is suggested if no a priori
information about the light curve shape is available as in our case.

First, for each target we searched for pulsed signals using the
whole available data span. Later, each observation block was
analyzed to check for possible flux and/or pulse profile
variability.

We also performed a preliminary maximum likelihood analysis (ALIKE task) on 
the \agile\ data for the regions containing our targets in order to exploit
the instrument's imaging capabilities to assess 
gamma-ray source parameters.
Here, we focus on timing analysis 
leaving detailed source positioning, flux and (phase-resolved) spectral analysis for future papers when higher counts photon statistics on each target will be available and instrument effective area calibrations will be consolidated.

\section{New gamma-ray pulsars}
\begin{figure}
\centering
\resizebox{\hsize}{!}{\includegraphics[angle=00]{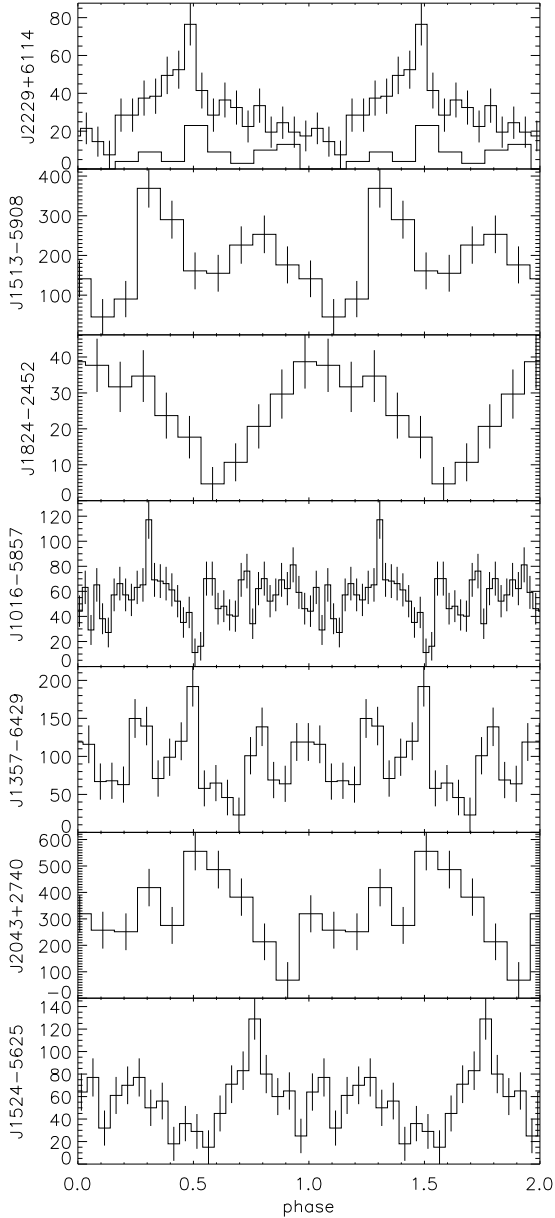}}
\caption{\label{all} Background-subtracted $E>100$ MeV folded pulse profiles 
(except for the $E>50$ MeV J1357--6429 light curve) of the pulsars shown in Table \ref{newpulsars}.
Events extraction radius is optimized as a function of energy (see the text).
All pulse profiles are obtained using G event class only except for PSR\,J1513+5908 and PSR\,J2043+2740 that have higher counts statistics because detections are obtained including G+L event class. For PSR\,J2229+6114 the histogram of the $\sim$75 (G+L) events with $E>1$ GeV is also shown. Absolute timing is performed for each target:
the main radio peak (1.4 GHz) corresponds to phase 0.
Possible fluctuations of the dispersion measure over the considered time interval
are not expected to significantly affect phasing results, given the time resolution
of the available gamma-ray light curves.}
\end{figure}

\begin{figure}
\resizebox{\hsize}{!}{\includegraphics[angle=00]{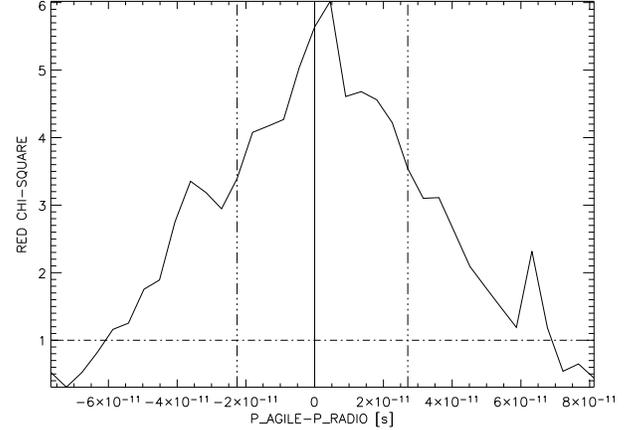}}
\caption{\label{J2229rc} PSR\,J2229+6114: gamma-ray period search result 
(period trials versus $\chi^2$ Pearson statistics). The radio period (vertical line at 
$P_{\rm{AGILE}}-P_{\rm{RADIO}}=0$) is 51.64101208(3) ms (PEPOCH = 54575.5856 MJD). The most 
significant gamma-ray pulse profile is obtained for $P_{\rm{AGILE}}^{Best}=51.641012085(2)$,
a value within the 1$\sigma$ error on the radio period (vertical dashed lines).}
\end{figure}

\begin{figure}
\resizebox{\hsize}{!}{\includegraphics[angle=00]{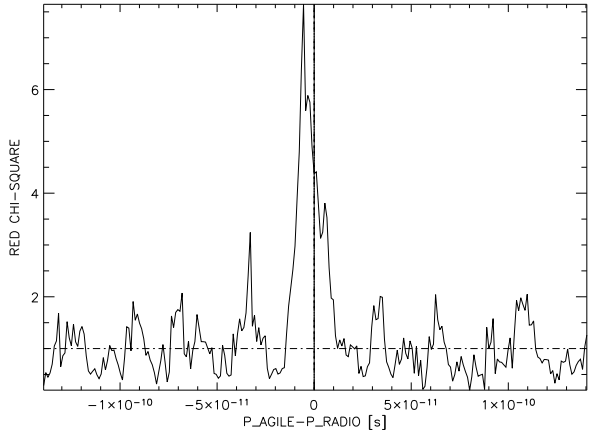}}
\caption{\label{J1824xxx} PSR\,J1824--2452: gamma-ray period search result. 
The radio period is 3.0543151208713(1) ms (PEPOCH = 51468 MJD). The most significant 
gamma-ray pulse profile is obtained for $P_{\rm{AGILE}}^{Best}=3.05431511(1)$ ms. The 
corresponding radio--gamma periods discrepancy  $\lesssim$10$^{-11}$ s is comparable to
the period search resolution in the considered data span (54339--54344 MJD).} 
\end{figure}

Table \ref{newpulsars} lists the emission parameters of the seven
pulsars discussed in this Letter.
The resulting radio-aligned light curves are plotted in Figure \ref{all}
where for each pulsar the actual bin size and energy range has been adjusted
according to the available statistics and light-curve structure. 
We note that in all cases radio and gamma-ray timing
results are compatible, with the highest significance frequency
detected in gamma rays within the 
errors of the radio 
ephemeris value, considering also the period search resolution.
Examples of the exploration of much larger frequency search
grids are shown in Figures \ref{J2229rc} and \ref{J1824xxx}
for PSR\,J2229+6114, for which the best gamma signal is within
1$\sigma$ from the radio peak, and for PSR\,J1824--2452,
for which the radio--gamma frequency discrepancy is
comparable with the gamma-ray period search resolution.
Both $\chi^2$ statistics and $Z^2_n$/$H$-test
provide comparable detection significances, except for PSR\,J2229+6114.
The $Z^2_n$ test applied to this pulsar provides  slighly better results than the
other statistics ($Z^2_1=39.5$, $Z^2_2=45.9$ corresponding to an $\sim$6$\sigma$ detection).
Furthermore, we verified that our analysis procedure (potentially affected by instrument-related systematic errors and
biases in events extraction criteria) does not produce fake
detections at a significance level above 3$\sigma$ when the
radio-ephemeris are applied to randomly extracted \agile\
data.


Four targets are also firmly detected by the likelihood spatial
analysis. It is not surprising  that spatial detection be missing
for weaker targets at the present \agile\ exposure level. In
high-background regions of the Galactic plane, our pulsed emission
search sensitivity can be better than our spatial analysis
sensitivity. Furthermore, timing analysis can be applied to the
full \agile\ event list (i.e. G+L event classes extending up to
60\degr\ from the center of the field of view), while, currently,
spatial analysis is only fully calibrated for the G class events
detected within 40\degr\ from the center of the field of view.

However, some of the non-detections in spatial analysis are, indeed, puzzling.
According to Figure \ref{all}, PSR\,J2043+2740 has a pulsed flux and light curve very similar to PSR\,J1513--5908 (note that both are based on G+L events). 
Moreover, J2043+2740 is out of the Galactic plane, in a region well exposed and with low 
diffuse emission. Thus, it should have been detected by the image analysis more easily 
than J1513--5908, which sits in a higher background region, unless, of course, they have very different spectra.


Pulsars in Table \ref{newpulsars} are ranked according to their
overall detection significance.
Pulsars firmly detected both through
timing and spatial analyses are placed above the line.
Pulsars detected through timing analysis alone await
longer simultaneous gamma-ray and radio observations for confirmation
and reliable luminosity estimation.

In Table \ref{newpulsars}, the most significant detection is
PSR\,J2229+6114 for which \agile\ detected pulsed emission
(radio/gamma-ray periods discrepancy $\lesssim$$10^{-11}$ s,
Figure \ref{J2229rc}) and pinpointed the most likely position
($l=106\fdg86$, $b=2\fdg94$)
to $\sim$$0\fdg2$ from the radio pulsar. Our detection provides a
reliable identification for the previously unidentified EGRET source
3EG\,2227+6122 ($l=106\fdg53$, $b=3\fdg18$).  The \agile\ source position,
pulsed flux, and photon
index ($\sim$2.2) are consistent with the EGRET values
\citep{hartman99}. The gamma-ray light curve of this pulsar (detected
up to over 1 GeV),
featuring just one prominent peak shifted $\sim$ 180\degr\ in phase from the
radio main peak, is shown in Figure \ref{all}.
It is worth noting that assuming a distance of $\sim$ 3 kpc
inferred by X-ray observations \citep{halpern01}
and isotropic emission, the pulsar gamma-ray efficiency would be $\sim$ 0.5,
a factor of 20 higher than that quoted in Table \ref{newpulsars}.
In modern pulsar beaming models (in particular high-altitude
models) the assumption of isotropy in luminosity calculation could be a better approximation in most cases, 
implying that efficiencies should be increased by roughly an order of magnitude from those
based on the commonly assumed 1 sr beam \citep{watters09}.
Anyway, precise efficiency measurements await for a better assessment
of pulsar distance, flux, and beam geometry.

PSR\,J1513--5908 (B1509--58) was detected by COMPTEL in the 1-10 MeV range, while
EGRET reported \citep{fierro96} only marginal evidence for a weak
$<$4$\sigma$ source at $\sim$ 1\degr\ from the radio position,
with a pulsed emission upper limit
 of $<$$58\times10^{-8}$ ph cm$^{-2}$ s$^{-1}$.
The \agile\ discovery of  pulsed emission from PSR\,B1509--58 relies
also on a 6.4$\sigma$ detection of a gamma-ray source (possibly multiple sources) at
$\lesssim$$0\fdg4$ from the radio position.
The main gamma-ray peak at phase $\sim$ 0.35 in the \agile\
light curve
is aligned with the soft gamma-ray peak seen by COMPTEL
in the 0.75--30 MeV band ($0.35\pm0.03$; \citealt{kuiper99}) slightly trailing
the hard X-rays single peaked light curves \citep{saito97,rots98}.
A second possibile peak in the \agile\ light-curve at $\sim$ 0.85
could have correspondence with the marginal feature seen in the 10--30 MeV
COMPTEL light curve. 
The COMPTEL sharp spectral break between 10 and 30 MeV
\citep{kuiper99} is confirmed by our \agile\ point at 100 MeV.
Our data imply a softening of the photon index from $\sim$ 1.7 to
$\sim$ 2.5 going from tens to hundreds of MeV. Such a low-energy
break, compared to the more common GeV spectral break in gamma-ray
pulsars, could be the telltale signature of the photon splitting
process inducing significant e.m. cascade attenuation due to the
strong magnetic polar field (greater than $10^{13}$ G) of this pulsar
\citep[e.g.][]{harding97}.

At variance with all the other targets, the millisecond pulsar
J1824--2452 in the Globular Cluster M28 was detected by \agile\,
with good significance (greater than 4$\sigma$) and perfect radio--gamma periods match, 
only in the time interval
54339--54344 MJD (Figure \ref{J1824xxx}).
The main radio peak at 1.4 GHz is coincident with the broad
single peak seen in gamma rays.
Only marginal
detection was obtained integrating other observations with
comparable exposure or the whole data span. Noise fluctuations
could possibly explain the apparent variability.
Alternatively, although its gamma-ray
efficiency and high stability of spin parameters are compatible
with rotation-powered emission, some additional mechanism
disturbing the neutron star magnetosphere in the dense cluster
environment could be invoked to explain the variable gamma-ray
phenomenology of this peculiar pulsar. \agile\ timing failures at
submillisecond level in some observations (mimicking source
variability) cannot be excluded. However, this seems unlikely,
since we verified both timing accuracy and stability at $\sim$ 200
$\mu$s level with Vela pulsar observations. Confirmation of this
tantalizing result about physical variability will rest on longer monitoring campaigns.

PSR\,J1016--5857 stands out in Table \ref{newpulsars} for its very high
efficiency in converting rotational energy loss into high-energy
gamma rays. This may be ascribed to distance uncertainties, as
happened to be the case for the recently discovered PSR\,
J2021+3651 \citep{halpern08} for which 
the distance derived from the dispersion measure is
certainly overestimated. Furthermore, we note that the position
($l=284\fdg47$, $b=-0\fdg94$) and flux of the \agile\
gamma-ray source is only marginally compatible with
3EG\,1013--5915,  for which multiple associations were proposed
\citep{hartman99}. The region, originally covered by the
gamma-ray source 2CG\,284--00, discovered by COS-B \citep{swan81}, is very complex.
Other nearby sources, and in particular pulsar PSR\,J1016--5819 (not belonging
to our sample), could significantly contribute
to the high flux observed by \agile\ in this region.

Of the remaining three pulsars, detected only through timing
analysis, the most notable is certainly PSR\,J2043+2740 which,
with an age in excess of one million years, would be the oldest
ordinary (nonrecycled) pulsar seen in the gamma-ray domain. The
precise measurements of its flux will yield its luminosity and
thus its efficiency, a parameter of paramount importance for the
understanding of gamma-ray emission as a function of the pulsar
age.


According to their timing and spatial analysis outcomes, light-curve profiles, and
radio--gamma phasing upshots, PSR\,J2229+6114, J1513--5908 and J1824--2452
can be considered rather solid detections.
If the detection of J1016--5857, J2043+2740, J1357--6429, and J1524--5625 will
also be firmly established, our seven object
sample, together with the detection of J2021+3651 \citep{halpern08},
would imply that gamma-ray emission is a common feature of
luminous radio-loud pulsars, be they young or old. Indeed, our list
encompasses the second youngest (J1513--5908) and by far the
oldest nonrecycled pulsar (J2043+2740) detected at gamma-ray energies.
The gamma-ray pulsar search and the improvement of \agile\ timing 
procedures requiring unprecedented long integrations is an ongoing effort, 
in close collaboration with the
radio observers. We trust that the new gamma-ray pulsar detections
presented here will trigger multiwavelength observations (e.g. in
X-rays) simultaneous to the continuously improving \agile\ and
\emph{Fermi} (formerly \emph{GLAST}) exposures on these targets.
The availability of high-resolution light curves and their
precise multiwavelength phasing will pave the way to improve
pulsar models as well as our understanding of the pulsar
population as a whole.\smallskip

We thank the referee for her/his useful recommendations and comments. The Parkes radio
telescope is part of the Australia Telescope,
which is funded by the Commonwealth of Australia for operations as
a National Facility managed by CSIRO.
We thank Jules Halpern and Fernando Camilo for useful discussions and contributions
on gamma-ray pulsar timing.

\end{document}